\documentclass[journal,compsoc]{IEEEtran}

\usepackage{color}
\usepackage{graphicx,epstopdf}
\usepackage{array}
\usepackage{url}
\usepackage{cite}
\usepackage[utf8]{inputenc}
\usepackage{ragged2e}
\usepackage{amsmath}
\usepackage{amssymb}
\usepackage{algorithm}
\usepackage{algpseudocode}
\usepackage{enumitem}

\fontfamily{pcr}\selectfont

\begin{document}

\title{IBAC: An Intelligent Dynamic Bandwidth Channel Access Avoiding Outside Warning Range Problem}


\author{Raja Karmakar,~\IEEEmembership{Member,~IEEE}, and~Georges~Kaddoum,~\IEEEmembership{Senior~Member,~IEEE}

\thanks{R. Karmakar and G. Kaddoum are with Department of Electrical Engineering, ETS, University of Quebec, Montreal, Canada (Email: raja.karmakar.1@ens.etsmtl.ca, georges.kaddoum@etsmtl.ca).}%
} 


\IEEEtitleabstractindextext{
\begin{abstract}

IEEE 802.11ax uses the concept of primary and secondary channels, leading to the {\em Dynamic Bandwidth Channel Access (DBCA)} mechanism. By applying DBCA, a wireless station can select a wider channel bandwidth, such as $40/80/160$ MHz, by applying the channel bonding feature. 
However, during channel bonding, inappropriate bandwidth selection can cause collisions. Therefore, to avoid collisions, a well-developed media access control (MAC) protocol is crucial to effectively utilize the channel bonding mechanism. In this paper, we address a collision scenario, called {\em Outside Warning Range Problem (OWRP)}, that may occur during DBCA when a wireless station interferes with another wireless station after channel bonding is performed. Therefore, we propose a MAC layer mechanism, {\em Intelligent Bonding Avoiding Collision (IBAC)}, that adapts the channel bonding level in DBCA in order to avoid the OWRP. We first design a theoretical model based on Markov chains for DBCA while avoiding the OWRP. Based on this model, we design a {\em Thompson sampling} based Bayesian approach to select the best possible channel bonding level intelligently. We analyze the performance of the IBAC through simulations where it is observed that, comparing to other competing mechanisms, the proposed approach can enhance the network performance significantly while avoiding the OWRP.

\end{abstract}
}

\maketitle

\IEEEdisplaynontitleabstractindextext

\IEEEpeerreviewmaketitle

\section{Introduction}
\label{sec:intro}

The growing developments of smart cities across the world have increased the need for high speed wireless access networks over IEEE 802.11 technologies. In this direction, the IEEE 802.11ax~\cite{bellalta2016ieee}, commonly known as {\em High Efficiency Wireless Local Area Networks} (HE-WLANs), is the successor of IEEE 802.11ac~\cite{standardDot11ac}, known as {\em High Throughput WLANs} (HT-WLANs), to support data rates around $10$ Gbps over the $5$ GHz channel. Basically, as per legacy IEEE 802.11 standards, the width of a channel is $20$ MHz. The IEEE 802.11ac enhances the physical (PHY) layer by introducing the advanced channel bonding mechanism~\cite{bejarano2013ieee}, where several consecutive $20$ MHz channels can be combined together to generate wider channels ($40/80/160$ MHz). However, a station may not always require the full capacity or bandwidth of the wider channel. Consequently, the concept of {\em primary} and {\em secondary channels}~\cite{bejarano2013ieee} are introduced in IEEE 802.11ac, where a station can acquire a $20$, $40$, or $80$ MHz channel as the primary channel and an extension of the same bandwidth can be acquired as the secondary channel. 

\begin{figure}[!t]
\centering
\includegraphics[width=\linewidth]{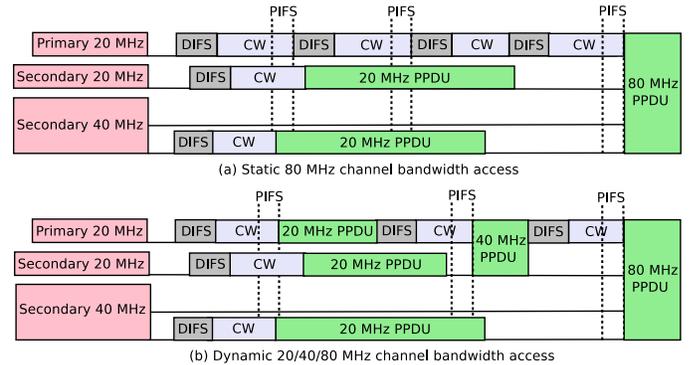}
\caption{Static and dynamic bandwidth channel access.}
\label{fig:dbca}
\end{figure}
\textbf{Dynamic Bandwidth Channel Access:}
For efficient utilization of the primary and seconday channels, {\em Dynamic Bandwidth Channel Access (DBCA)}~\cite{park2011ieee} has been introduced at the media access control (MAC) sublayer in IEEE 802.11ac. DBCA works over the distributed coordination function (DCF) of IEEE 802.11 to use the primary and secondary channels. In DBCA, a binary exponential based backoff algorithm, which is similar to the DCF, is used by a station to access the primary channel. A contention window (CW) is used by the station to maintain the backoff time interval. Before transmission of a packet, the backoff time is chosen uniformly in the $(0,...,W-1)$ range, where $W$, known as the CW, depends on the transmission history of the packet. At the first transmission attempt, $W$ is set to $W=CW_{min}$ which is defined as the minimum contention window. Each value of $W$ leads to a backoff stage which denotes a time interval of waiting for a station before data transmission. When a transmission is unsuccessful, $W$ is doubled. The maximum value of $W$ ($CW_{max}$) is $CW_{max}=2^m CW_{min}$, where $m$ represents the maximum backoff stage. 

Before data transmission, a station senses the primary channel over a time interval known as DCF interframe space (DIFS). After the DIFS interval, the station chooses a CW and waits for the number of slots defined by this CW. 
After the DIFS and CW, if the station senses the primary channel as idle, it proceeds to sense the secondary channel. 
If the secondary channel is free, the station transmits data by using both the primary and the secondary channels. Otherwise, data is transmitted only over the primary channel. 
Therefore, the channel bonding-based DBCA mechanism can increase the system performance~\cite{park2011ieee}.
Fig.~\ref{fig:dbca}(a) shows the access of the static $80$ MHz of channel width, where a station transmits a physical layer convergence procedure protocol data unit (PPDU) on an $80$ MHz of channel bandwidth. Here, the primary channel is idle for the DIFS plus backoff time and the three consecutive secondary channels are found to be idle for the point coordination function interframe space (PIFS) immediately after the expiration of the CW period. In this case, the station cannot transmit the data until all the secondary channels are free. However, in DBCA, the station can use $20$, $40$, or $80$ MHz of channel bandwidth for data transmission depending on the availability of the secondary channels, as illustrated in Fig.~\ref{fig:dbca}(b).

\begin{figure}[!t]
\centering
\includegraphics[width=\linewidth]{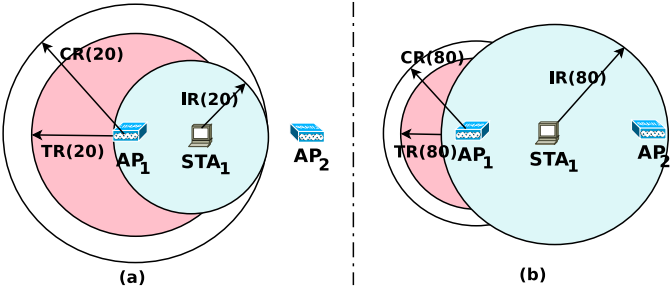}
\caption{OWRP: (a) The AP $AP_1$ and station $STA_1$ use $20$ MHz channel for exchanging RTS/CTS. The size of Interference Range (IR) ($20$ MHz) is less than the size of Transmission Range (TR) ($20$ MHz). In this case, the AP $AP_2$ is situated outside of IR ($20$ MHz) of station $STA_1$. (b) When $AP_1$ transmits data by using $80$ MHz channel, $AP_2$ will become inside of IR ($80$ MHz) of $STA_1$ and thus, collisions may occur.}
\label{fig:owrp}
\end{figure}

\textbf{Outside Warning Range Problem:}
The authors in~\cite{park2011ieee,Xhafa2008RTS,Jang2015RTS,ChangDBCAperformance} reveal that {\em Request to Send (RTS)/Clear to Send (CTS)} handshake mechanisms of the IEEE 802.11~\cite{standardDot11} cannot perform well when the channel bonding approach is adopted. For instance, the {\em hidden channel problem}~\cite{hiddenNode2011} can arise in the network when two access points (APs) attempt to send data over overlapped bonded channels. This limitation of the RTS/CTS scheme leads to the so-called {\em Outside Warning Range Problem (OWRP)} in multi-hop wireless networks. 
To overcome the hidden channel problem, the station that receives a RTS frame responds with a CTS frame.
However, when channel bonding is applied, collisions may happen in the channel even if the interfering station is located outside the transmission range of the receiver. This is because the Interference Range (IR) is increased due to the enhancement of the channel width after channel bonding~\cite{XuTR:2002} and~\cite{DBS:2018}. This phenomenon is known as {\em OWRP}, as illustrated in Fig.~\ref{fig:owrp}. The enhancement of the channel width reduces the gap between two consecutive channels, and also increases the channel overlapping probability. As a result, the IR is increased. 
When a RTS frame is transmitted to station $STA_1$ from AP $AP_1$ through a $20$ MHz channel, an IR~\cite{XuTR:2002} will be induced at station $STA_1$, as shown in Fig.~\ref{fig:owrp}(a). At this time, AP $AP_2$ is not situated within the transmission range of the $20$ MHz channel bandwidth or in the IR of station $STA_1$. Here, the Carrier Sensing Range (CR) is defined as the range within which a station can sense the power generated by the transmitter. For a given transmission power, as the channel bandwidth increases, the Transmission Range (TR)~\cite{XuTR:2002} decreases but the size of the IR increases~\cite{XuTR:2002} and~\cite{DBS:2018}. For example, if the data is transmitted through $80$ MHz channel using a channel bonding mechanism, as shown in Fig.~\ref{fig:owrp}(b), the IR of station $STA_1$ is increased when the transmission is performed using channel bonding. As a result, $AP_2$ will remain outside of the TR~\cite{XuTR:2002} of $AP_1$, which is due to the use of $80$ MHz channel; however, $AP_2$ will fall inside the IR of station $STA_1$. Therefore, a collision will occur when the transmission of $AP_1$ overlaps with that of $AP_2$. 

\subsection{Existing Approaches and Their Limitations}
Several works have been proposed in the open technical literature to address DBCA~\cite{park2011ieee,Saketh2017,Ayush2017,ChangDBCAperformance,Mun-Suk:2017,Stelter2014,KimDBCA2017,ByeonDBCA2015,DBS:2018}. A two dimensional Markov chain-based model is designed in~\cite{Saketh2017} to analyze the impact of the DBCA approach over the standard use of primary and secondary channels. This work also focuses on the unfairness of channel access during channel bonding. The work in~\cite{Ayush2017} highlights how unfairness is increased during channel access in DBCA. 
Unfairness has been proven to reduce the average throughput of a wireless station, therefore the authors in~\cite{Ayush2017} propose an adaptive resource reservation mechanism to support fair channel access in DBCA. 
However, the works~\cite{Saketh2017,Ayush2017} do not overcome the OWRP in their Markov chain models.
Authors in~\cite{park2011ieee} discuss the static channel access scheme for the case of $40$ MHz channels and extend this approach to the dynamic access of $80$ MHz channels, based on clear channel assessment (CCA). The presented results show that a wider channel is effective in a dense networking environment. In~\cite{ChangDBCAperformance}, an analysis is presented showing the performance of IEEE 802.11ac DCF in the presence of hidden nodes by considering overlapping basic service set (BSS). This work highlights that the shortcomings of the traditional RTS/CTS handshaking schemes need to be modified before utilizing in the IEEE 802.11ac standard. 
The problem of low throughput in networks containing a large number of stations is due to the random backoff in the carrier sense multiple access with collision avoidance (CSMA/CA) approach. The work in~\cite{Liu:2016} addresses this problem by designing a traffic-aware backoff functionality. 

By using a static channel access approach, the work in~\cite{Mun-Suk:2017} estimates the throughput of IEEE 802.11ac networks by developing a Markov chain model that describes the backoff mechanism in IEEE 802.11ac. 
A dynamic channel width access mechanism using a virtual reservation approach of the primary channel is developed in~\cite{KimDBCA2017}. This approach attempts to reduce the performance of bottleneck and scalability problems. Stelter \textit{et al.}~\cite{Stelter2014} designed a dynamic channel access mechanism for IEEE 802.11ac-based networks by employing the CCA functionality. Authors in~\cite{ByeonDBCA2015} investigate the performance of the dynamic bandwidth channel access operation in IEEE 802.11ac. In this work, an algorithm is also designed to enable/disable the dynamic channel bandwidth operation in an adaptive way. Considering the IEEE 802.11n standard, a MAC layer scheme that reduces the probability of collision in channel access without exploring DBCA in IEEE 802.11ac is proposed in~\cite{GMAC2013}. 

For IEEE 802.11ax, the work in~\cite{han2020deep} proposes a probabilistic channel aggregation mechanism to maximize the network throughput. Here, a user intelligently aggregates a secondary channel using a probability considering the network traffic of the secondary channel. To maximize the energy efficiency of the APs, a deep reinforcement learning based scheme that controls the channel bonding and transmit power of APs in an energy-efficient way is proposed in~\cite{luo2021energy}. In~\cite{barrachina2019dynamic}, by using continuous-time Markov networks, the behavior of different dynamic channel bonding policies is examined in spatially distributed dense networks. The authors in~\cite{nabil2020stochastic} discuss the optimal utilization of available frequency bands for channel bonding, considering channel acquisition demands of APs. An analytical framework is proposed in~\cite{lanante2020analysis} to model the channel bonding mechanism as a function of PHY and MAC parameters, and a channel bonding algorithm is described considering the overlapping of co-channel interference. The work~\cite{han2019capacity} analytically studies the performance of a multi-channel bonding mechanism that supports delay-sensitive multimedia applications. The work in~\cite{barrachina2021wi} presents a system to simultaneously measure all Wi-Fi channels that support channel bonding in the 5 GHz band with a microsecond scale granularity. The authors in~\cite{barrachina2019online} propose an online decentralized primary channel selection approach for dynamic channel bonding, that improves the network throughput by considering both the availability of the primary channel and the pursuit of the secondary channels. To avoid hidden channels, the mechanism proposed in~\cite{karmakar2020smartbond} predicts the channel bandwidth for channel bonding, that can reduce the interference. However, this work does not explicitly consider the OWRP.

The work in~\cite{yang2019hiatus} adjusts the channel width dynamically based on the packet error rate to avoid hidden channel interference on the secondary channel.
Jang \textit{et al.}~\cite{jang2017post} address the hidden channel problem by using a CCA procedure which is performed after transmission is completed. Moreover, based on the CCA, the channel bandwidth is adapted dynamically.
The very recent work in~\cite{park2020widercast} enables channel bonding operation for multicast transmission by using a bandwidth signaling mechanism, in which multiple stations deliver their supportable channel bandwidth information simultaneously to the AP. All these approaches use different heuristics to detect the effect of the hidden channel interference and do not explicitly consider the OWRP issue during the selection of the secondary channel.
To handle DBCA, the concept of OWRP is first addressed in~\cite{chen2017channel} and~\cite{DBS:2018}. By analyzing the relationships among the TR, CR and IR, the work in~\cite{chen2017channel} proposes a threshold-based MAC protocol to avoid the OWRP in IEEE 802.11ac networks. The work in~\cite{DBS:2018} is basically an extension of~\cite{chen2017channel} and the authors in~\cite{DBS:2018} design a MAC layer protocol for IEEE 802.11ac-based networks to avoid collisions due to the carrier sensing decreasing problem (CSDP) and OWRP. However, the mechanism proposed in~\cite{DBS:2018} is also threshold-based, where its performance varies depending on the network environment. 


Hence, apart from~\cite{chen2017channel} and~\cite{DBS:2018}, existing works do not handle the OWRP issue when using a higher channel bandwidth during data transmission; whereas,~\cite{chen2017channel} and~\cite{DBS:2018} lack adaptability to different network scenarios due to their threshold-based mechanisms. However, when the channel bonding feature is utilized, the OWRP can seriously affect the network performance as the number of stations increases. 
Moreover, from Fig.~\ref{fig:owrp}, it can be noted that the MAC protection mechanism defined in IEEE 802.11 cannot avoid collisions due to the OWRP, which results in a degradation of the throughput of HE-WLANs. Therefore, handling the OWRP still remains an open research challenge that needs to be tackled dynamically by using an online learning approach so that the stations can adjust channel width intelligently in order to avoid the OWRP in the network. 



\subsection{Pilot Study}
\label{sec:motivation}

We have carried out a pilot experiment to find out the impact of OWRP on the system performance in IEEE 802.11ac. The details of the experimental setup are provided in Section~\ref{sec:performance}. Here, we consider two APs which form two BSSes. Since it was shown in Fig.~\ref{fig:owrp} that the distance between two APs can influence the OWRP significantly, we have considered different distances between the two APs in our experiment. We have a set of $d$ values of distances between APs (DAP), where $d=[80, 100, 120, 140, 160]$. In each BSS, there are $5$ STAs whose positions are selected randomly. 

\begin{figure}[!t]
\centering
\includegraphics[scale=0.23]{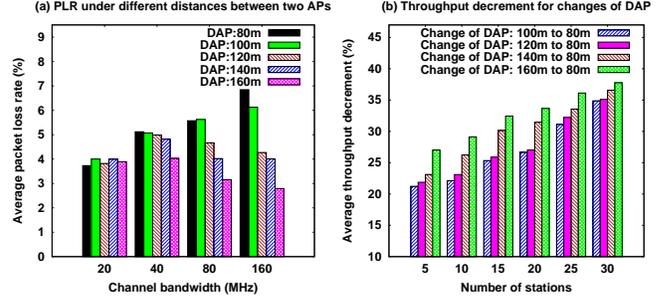}
\caption{(a) PLR variations for varying distances between two APs under different channel bandwidths and (b) Decrement in throughput for DAP changes.}
\label{fig:motivation}
\end{figure}

\noindent{\textbf{(i) Average packet loss rate:}}
Fig.~\ref{fig:motivation}(a) illustrates the variations in the average value of Packet Loss Rate (PLR) considering several distances between the APs and different values of channel bandwidth. From Fig.~\ref{fig:motivation}(a), it can be noted that the PLR and DAP are inversely proportional. The reason for this is that the probability of covering an AP by the IR of STA is reduced as the DAP increases, and thus the effect of collision due to OWRP is decreased. In addition, higher channel bandwidth results in an increase of the IR of STA. As a result, the impact of the OWRP also increases, which leads to more packet collisions in the network and consequently, the rate of packet loss in the network is increased. 
For instance, when the channel width is $40$ MHz, the PLR in $80$m DAP is increased approximately by $1.77$ times as compared to the PLR in $160$m DAP. When the channel width is increased from $20$ MHz to $80$ MHz, and if the DAP is $80$m, the PLR increases approximately by $49.3\%$; However, for the same change in the channel width, if the DAP is $160$m, the PLR decreases approximately by $18.97\%$. This is because increasing the DAP decreases the probability of OWRP, and hence increasing the channel width helps boost network performance in this instance.

\noindent{\textbf{(ii) Average rate of throughput decrement due to the decrease of DAP:}}
From Fig.~\ref{fig:motivation}(b), it can be observed that, for any $k$ stations, with $k=[5, 10, 15, 20, 25, 30]$, the rate of throughput decrement is the highest when the DAP is reduced to $80$m from $160$m. As the DAP decreases, the chances of getting inside of the coverage area of the IR of an STA become higher for an AP. As a result, for the available set of DAP, the effect of the OWRP is most significant for a DAP of $80$m and this effect becomes least significant for a DAP of $160$m. In addition, the probability of packet drop is higher in a congested network, which results in a lower throughput in the network; this degradation of throughput is increased as the DAP decreases, as illustrated in Fig.~\ref{fig:motivation}(b). For a decrement in DAP from $160$m to $80$m, when the number of stations is increased from $5$ to $30$, the increase in throughput decrement is approximately $39.7\%$.

As a result, it is critical to design a model that adapts the channel bonding level during DBCA to avoid OWRP and take use of the benefits of employing a higher channel width, while considering the current network situation. 

\subsection{Our Approach}
To address the aforesaid problems, in this paper, we propose a MAC layer mechanism, {\em Intelligent Bonding Avoiding Collision (IBAC)}, that intelligently selects the channel bonding level in DBCA while avoiding the OWRP. Since the RTS/CTS scheme cannot detect the OWRP properly, we apply a probabilistic approach in IBAC to avoid the OWRP while accessing the secondary channel. In this direction, we first derive the estimated system throughput for DBCA by designing a theoretical model based on Markov chains. Then, on the basis of the estimated throughput, we apply the {\em Thompson sampling (TS)}~\cite{Thompson:1933} based Bayesian approach to find the best DBCA channel bonding level intelligently. TS has several advantages -- (a) it is a reinforcement learning algorithm which is suitable for problems demanding online decisions~\cite{wassermann2020adaptive,kafi2019line,joshi2008sara}, (b) it generates the probability distribution of the success rate of applying an action based on it knowledge of exploiting the action in the past, and (c) in order to maximize the cumulative rewards obtained by applying actions, the application of exploration and exploitation leads to the learning of a new environment and utilizing the best past knowledge as well. In TS, the actions are chosen sequentially by a manner such that a balance is maintained between decisions exploring new information to improve future performance and decisions exploiting the past knowledge to maximize present performance.

Signal-to-noise ratio (SNR) is a key parameter in finding the IR~\cite{XuTR:2002,DBS:2018}. Thus, we use the SNR to select the channel width.
TS is composed of three steps -- the calculations of {\em prior}, {\em likelihood}, and {\em posterior distributions}. Since the prior provides the best assessment of computing the probability of an event based on the present knowledge, we calculate the prior of selecting a channel bonding level based on the present SNR value of the channel. The likelihood finds the conditional probability of the occurrence of an event conditioned on the probability of another event. Thus, we employ the likelihood to compute the probability of achieving the throughput estimated by the proposed Markov chain-based model, by knowing the probability of selecting a channel bonding level. Following this, based on the prior and likelihood, the posterior distribution is applied for selecting the best possible channel bonding level. Therefore, the posterior probability distribution selects the channel bandwidth such that the probability of selecting the channel width is maximized based on the present channel condition and can achieve the system throughput which is estimated by the Markov chain-based model.

\textbf{Reason for applying TS-based Bayesian approach:}
The probability distribution is the right machinery for quantifying uncertainty. Prior facilitates the determination of the probability distribution of an uncertain quantity related to the occurrence of an event, without taking any evidence of its impact on the system performance. By considering the likelihood function, the TS algorithm computes the probability distribution of an event based on its past performance. Based on the prior and likelihood, the posterior finds the probability of exploitation. Therefore, the likelihood, prior, and posterior make a natural balance between exploration and exploitation, and therefore TS can dynamically adapt in an unknown environment, considering the uncertainty of the event such as OWRP.
Since HE-WLANs support several channel bandwidths, we need to explore all the available channel widths under different network conditions. Thus, the size of the past information will be increased as various channel widths are selected in different network scenarios, and consequently the search space will be increased. Therefore, a probabilistic approach can be helpful in choosing the best possible channel width based on the past and predicted future system performances. In this context, the TS-based Bayesian approach presents an efficient mechanism where the probability distribution of applying the channels can be considered based on the past performances of the selected channels and the channel condition. In addition, considering the exploration history, channel widths can be exploited to achieve the best performance with the help of the prior, likelihood, and posterior distributions. Therefore, in an unpredictable environment (such as the wireless environment), a TS-based Bayesian approach can be useful for the selection of the best level of channel bonding by considering the present channel condition. 

\subsection{Contribution of this Work}
In order to avoid the OWRP in HE-WLANs, we design IBAC integrating the proposed Markov chain and TS-based online learning models. To the best of our knowledge, this work is the first to design a Markov chain model to overcome the OWRP during dynamic bandwidth channel access and accordingly, apply an intelligent adjustment of channel bandwidth at the time of DBCA.
Based on the intelligent DBCA avoiding the OWRP, the IBAC can lead to the performance improvement of IEEE 802.11ax standard. 
The main contributions of this work are summarized as follows:
\begin{enumerate}
 \item We design a Markov chain model to overcome the OWRP in DBCA. 
The proposed Markov chain model is built upon Bianchi's model~\cite{bianchi2000performance}, which designs the DCF in legacy IEEE 802.11 standards. 
Based on Bianchi's model, our proposed Markov chain model designs the DBCA avoiding the OWRP in HE-WLANs. 
 \item We propose an intelligent adaptation of channel bonding level in DBCA such that the resultant channel width after channel bonding overcomes the OWRP.
 \item A thorough performance analysis of the proposed mechanism is performed by implementing a prototype of the IBAC in network simulator (NS) version NS-3.33~\cite{ns3}, where the results demonstrate that the IBAC significantly enhances the network performance compared to other competing mechanisms. 
\end{enumerate}
Differences between the proposed model and the Bianchi's model are mentioned as follows:
\begin{itemize}
 \item Bianchi's model only designs the DCF in legacy IEEE 802.11 standards. By extending the Bianchi's model, our proposed model designs the DBCA mechanism in HE-WLANs such as IEEE 802.11ax.
 \item Bianchi's Markov chain model does not handle the DBCA and OWRP. However, when DBCA is applied, our model particularly solves the OWRP issue.
 \item The channel bonding process is not taken into account in Bianchi's model, whereas our model considers it, and thus the estimated system throughput is based on the application of different levels of channel bandwidth.
 \item In Bianchi's model, the packet transmission probability does not depend on the channel bandwidth. Since our model takes channel bonding and DBCA into account, the channel bandwidth specified during transmission influences the packet transmission probability.
\end{itemize}

\subsection{Organization of this paper}
The remainder of this paper is organized as follows. 
Details on the IBAC system model, the description of the proposed Markov chain-based OWRP avoidance model, and the proposed TS-based learning mechanism are discussed in Section~\ref{sec:pre}. Section~\ref{sec:performance} analyzes the performance of the IBAC through simulations, and Section~\ref{sec:concl} concludes this paper.




\section{IBAC: System Model}
\label{sec:pre}


\begin{figure}[!t]
\centering
\includegraphics[width=0.9\linewidth]{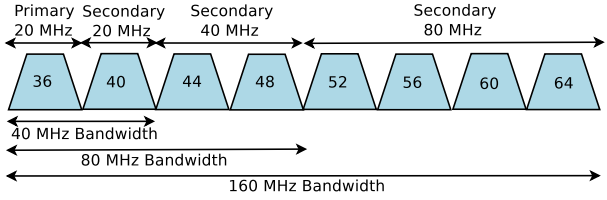}
\caption{Primary and secondary channel.}
\label{fig:cb}
\end{figure}

In this section, we detail the proposed IBAC aiming to tackle the challenge of avoiding the OWRP in the DBCA. 
At a given time in the data transmission, let a station use channel bonding level $c$. In our proposed Markov chain model, $c$ indicates that the station can use $2^{c-1}$ consecutive channels as the primary channel, where the effective channel bandwidth is $2^{c-1} \times 20$ MHz. The width of an individual channel is $20$ MHz. If $2^{c-1}$ channels are used as the primary channel, the next $2^{c-1}$ channels can also be utilized as a secondary channel. Let $u$ denote the maximum level of channel bonding, therefore a total of $2^{u-1}$ channels are available in the network. Moreover, let $\mathcal{M}$ be the set of available channel bonding levels, i.e. $|\mathcal{M}|=u$. In IEEE 802.11ax, there are a maximum of $4$ channel bonding levels and thus, $u=4$. 
Fig.~\ref{fig:cb} shows an example of channel bonding in IEEE 802.11ax. The numbers within the channels (trapezoids) denote the channel numbers in the frequency band of $5$ GHz.

\subsection{IBAC Architecture}

\begin{figure}[!t]
\centering
\includegraphics[width=\linewidth]{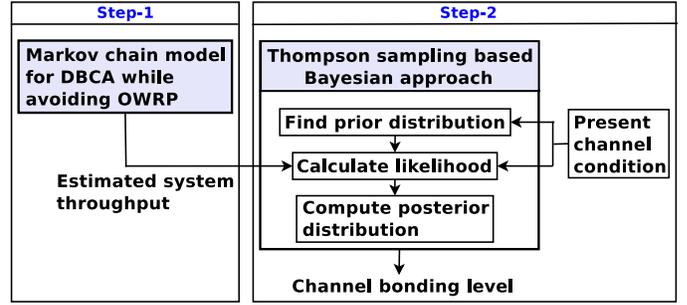}
\caption{IBAC architecture.}
\label{fig:dbca_owrp_model}
\end{figure}

Fig.~\ref{fig:dbca_owrp_model} illustrates the system architecture of the IBAC. We apply a probabilistic approach to avoid the OWRP while performing channel bonding for accessing the secondary channel. To design the IBAC, we employ the following steps.
\begin{enumerate}
 \item \textbf{Step-$1$ (detailed in Section~\ref{sec:Markov_model}):} We use a Markov chain to model the DBCA in order to overcome the OWRP, and derive the estimated system throughput in the network. 
 \item \textbf{Step-$2$ (detailed in Section~\ref{sec:TS}):} Based on the estimated throughput in Step-$1$, we apply a TS-based Bayesian approach to intelligently adapt the channel bonding level in DBCA.
\end{enumerate}

The greedy approach applies the concept of exploration and exploitation of the actions to maximize the reward. However, the greedy method has a tendency to explore the action space before exploiting the actions that were found to yield high rewards in the past experience. As a result, the greedy method can get stuck in the exploitation of a suboptimal action space. On the other hand, based on the Bayesian properties, TS can use uncertainty in applying the actions, where the probability distribution of the success rate of each action can be explored, and consequently the confidence of the action's outcome can be fully determined.

\textbf{Intersection of Step-$1$ and Step-$2$:} In Step-$1$, we use a Markov chain to model the DBCA in order to overcome the OWRP, and derive the estimated system throughput. In Step-$2$, based on the estimated throughput in Step-$1$, we apply a TS-based Bayesian approach to intelligently adapt the channel bonding level in the DBCA. In Step-$1$, the estimated normalized system throughput is expressed as the fraction of time duration where the wireless channel is used to transmit payload bits successfully, which is affected by the collisions in the channel. In case of DBCA, one of the major sources of collisions is the OWRP, and thus the system throughput is decreased as the OWRP increases in the network. Since the proposed Markov model estimates the normalized system throughput considering the OWRP, the estimated throughput can help tune the channel bonding level such that the impact of the OWRP can be avoided while using the DBCA in the network. Therefore, the estimated system throughput is considered in Step-$2$ to dynamically adapt the channel bandwidth in the DBCA.

\begin{figure}[!t]
\centering
\includegraphics[width=\linewidth]{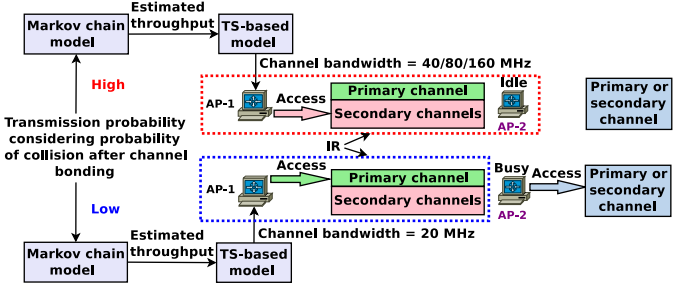}
\caption{Example of IBAC execution.}
\label{fig:IBAC_example}
\end{figure}

\textbf{Example of IBAC execution:}
An example of the execution of the IBAC is shown in Fig.~\ref{fig:IBAC_example} where two APs AP-1 and AP-2 are present in the network. The Markov chain-based model computes the transmission probability by determining the probability of collision after channel bonding in DBCA and then estimates the system throughput. As the probability of collision after channel bonding decreases, the probability of transmitting in higher channel widths increases, which implies a decrease in the probability of interfering with other APs or stations after an increase of the IR due to the channel bonding. Therefore, based on the proposed TS-based learning model, when the transmission probability is high, AP-1 selects a higher channel bandwidth ($40/80/160$ MHz). In this case, AP-2 is idle and it is inside the IR of AP-1, as illustrated in Fig.~\ref{fig:IBAC_example}. However, when the transmission probability is low, since AP-2 is now active and accessing a channel, AP-1 selects a channel width of $20$ MHz in order to reduce the IR induced by AP-1. 
Next, we present a detailed description of the proposed Markov chain-based model to overcome the OWRP in DBCA.

\subsection{Discrete Time Markov Chain Model for DBCA while Avoiding OWRP}
\label{sec:Markov_model}

In this paper, we consider a saturated network scenario, where all stations always have data for transmission and they participate in the channel access procedure. We use the term slot to refer to the time interval between two consecutive decrements of backoff time counter.


\subsubsection{Channel Access Model}

Let $p$ be the probability that a channel, randomly in a slot, is busy. Thus, the probability that no station transmits (i.e., the slot is free) is $(1-p)$. A station uses $c$ as the channel bonding level of the primary channel if $2^{c-1}$ consecutive channels are idle. After accessing the primary channel, the station utilizes the next $2^{c-1}$ consecutive channels as a secondary channel if they are free. Otherwise, only the primary channel is used for the transmission. Thus, the consecutive $2^{c-1}$ channels are utilized including both the primary and the secondary channels when the consecutive $2^{c-1}$ channels are idle. However, each station maintains a single CW that is used to access the primary channel only and the access to the secondary channel is independent of the CW. Therefore, if a station uses a channel bonding level $c$, it utilizes the first consecutive $2^{c-1}$ channels as the primary channel and the next $2^{c-1}$ channels as the secondary channel. For instance, if $c=3$, the station utilizes the first $80$ ($2^2\times 20$) MHz of bandwidth (first $4$ consecutive $20$ MHz channels) as the primary channel and, if idle, the next $80$ ($2^2\times 20$) MHz of bandwidth (next $4$ consecutive $20$ MHz channels) as the secondary channel. With the help of this model, we can manage the system under heterogeneous channel bonding environments, i.e., where different stations attempt to use different channel widths. 

Let $p_c$ be the probability of utilizing a channel bonding level $c$ by a station for its transmission. As each channel is independent, $p_c$ can be defined as
\begin{equation}\label{eq_p}
 p_{c} = \left\{ 
  \begin{array}{lr}
    (1-p)^{2^{c-1}}(1-p)^{2^{c-1}} & {\qquad{}\text{when $c < u$}}\\
    (1-p)^{2^{u-1}} & {\qquad{}\text{when $c = u$}}.
  \end{array}
  \right.
\end{equation}
Eqn.~(\ref{eq_p}) indicates that if the first $2^{c-1}$ channels and the consecutive $2^{c-1}$ channels are sensed as free, they will be used as the primary and secondary channels, respectively. At most, $2^{u-1}$ channels can be sensed.

\subsubsection{Collision and Transmission Models}

Let us consider that a station selects a level of channel bonding dynamically from $u$ available channel bonding levels. 
For a station transmitting data with a channel bonding level $c\le u$, the transmission would be successful if none of the $2^{c-1}$ channels experience a collision. 
We design our Markov chain model inspired from the virtual slot duration based model in~\cite{bianchi2000performance}, which proposes a Markov chain model only to design the DCF in legacy IEEE 802.11 standards. However, our proposed Markov chain model designs the DBCA avoiding the OWRP in high throughput WLANs, and therefore it can specifically handle the OWRP issue when channel bonding is applied.

When the level of channel bonding is $c$, a transmission would suffer a collision if a collision occurs in any of the available $2^{c-1}$ 
channels. Let $q$ represent the probability of collision in any randomly chosen channel. Moreover, let $q_c$ denote the probability of collision in any of the available $2^{c-1}$ channels. Thus, we have $q_{c} = 1-(1-q)^{2^{c-1}}$. Now, the interference range is directly proportional to the channel bandwidth, i.e. the range of interference increases as the channel bonding level increases. Hence, we can represent $q$ as $q \propto \frac{2^{c-1}}{2^{u-1}}$, i.e. $q=\kappa \big(\frac{2^{c-1}}{2^{u-1}}\big)$, where $\kappa$ is a proportionality constant. Therefore, we have 
\begin{equation}\label{eq_qc}
q_{c} = 1-\Big(1-\kappa \big(\frac{2^{c-1}}{2^{u-1}}\big)\Big)^{2^{c-1}}.
\end{equation}

\subsubsection{Collision Probability with OWRP}

The OWRP issue can arise in the network after a station senses the secondary channel as idle and starts its data transmission. Hence, in the OWRP, the previous slot is idle but the current slot becomes busy due to the activation of a station during the transmission of an arbitrary station. 
Let us consider that $\gamma$ represents the probability that a station accesses the channel after an idle slot. Considering the previous slot as idle, the probability that there is no other active station during the transmission of an arbitrary station is $(1-\gamma)^{n-1}$. Here, $n$ is total number of stations in the network. Let $r$ be the probability of collision given that the previous slot is idle and at least one station is currently active in the present slot during the transmission of an arbitrary station. Thus, in this case, $r$ can be defined as
\begin{equation}
r = (1-p)[1-(1-\gamma)^{n-1}].
\end{equation} 

Now, let $\Upsilon_C$ denote the probability of collision of an arbitrary station using channel bonding, given that the previous slot is empty but the current slot is busy (at least one station is active other than the arbitrary station). $\Upsilon_C$ can be represented as
$$\Upsilon_C = \frac{p_1q_1 + p_2q_2 + ... + p_{c}q_{c} + ... + (1-p)^{2^{u-1}}q_u}{r}$$
\begin{align}
 \label{tc}
    =\frac{p_1q_1 + p_2q_2 + ... + p_{c}q_{c} + ... + (1-p)^{2^{u-1}}q_u}{(1-p)[1-(1-\gamma)^{n-1}]},
\end{align}
where $p_{i}q_{i}$ is the probability of collision when a station applies the channel bonding level $i$, $i \in \{1,2,...,c\}$. In~(\ref{tc}), the numerator represents the probability of collision when a station utilizes a level of channel bonding. Therefore, after dividing~(\ref{tc}) by $r$, we obtain the probability of collision of an arbitrary station using channel bonding, given that the previous slot is empty but the current slot is busy.
\subsubsection{Markov Model for DBCA}

\begin{figure}[!t]
    \centering
    \includegraphics[width=\linewidth]{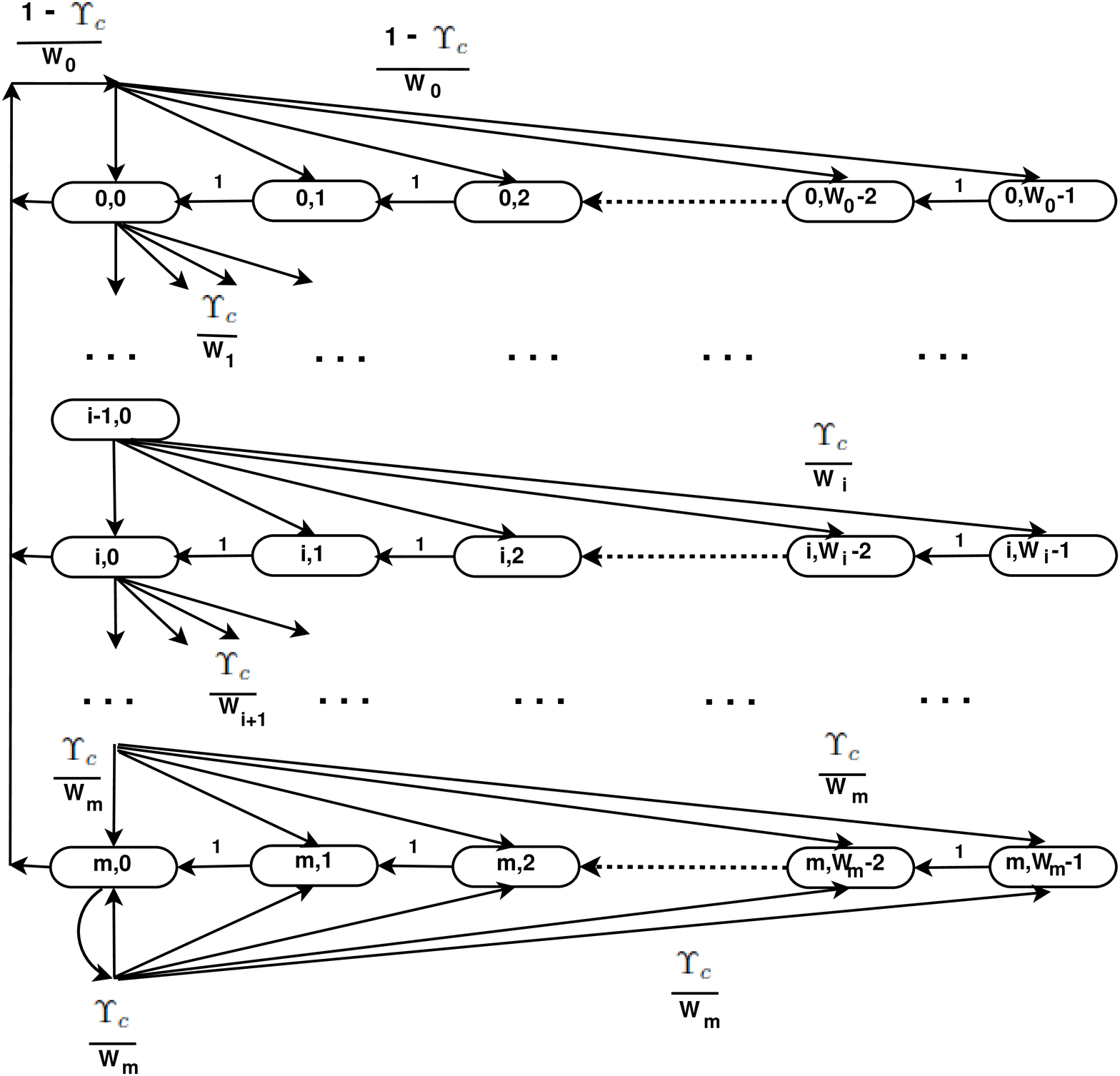}
    \caption{Markov chain model for DBCA.}
    \label{fig-model}
\end{figure}

Let $b(t)$ denote the stochastic process that represents the backoff time counter for a station at time $t$. Here, the beginning time instants of two consecutive slots are represented by $t$ and $t+1$, respectively. Moreover, the backoff time counter is decremented at the start of each time slot, where the decrement is stopped when the channel is found to be busy. In general, to represent a contention window, we consider the notation $W_i=2^i CW_{min}$, $i\in (0,m)$ denotes ``backoff stage". At time $t$, assume that $s(t)$ denotes the stochastic process representing the backoff stage $(0,...,m)$ of a station. The proposed Markov model which is a two-dimensional Markov chain is shown in Fig.~\ref{fig-model}.
We use a two dimensional Markov chain because we need two parameters, namely the backoff time counter and the stochastic process representing the backoff stage, to represent a state in the Markov chain. 
Thus, a state is represented by two tuples $(s(t),b(t))$. At each back-off stage, every station uses the average collision probability ($\Upsilon_C$) and average success probability ($1-\Upsilon_C$) for a transmission. 


\subsubsection{Packet Transmission Probability} 

In the Markov chain shown in Fig.~\ref{fig-model}, the only non null one-step transition probabilities are given as

\begin{equation}
\label{eq1}
     \left\{ {
        \begin{array}{lll}
        & P \{i,k | i,k+1\} = 1  & \hbox{$ k \in (0,W_i-2), i \in (0,m)$} \\
        & P \{0,k | i,0\} = \frac{1-\Upsilon_C}{W_0}  & \hbox{$ k \in (0,W_0-1), i \in (0,m)$} \\
        & P \{i,k | i-1,0\} = \frac{\Upsilon_C}{W_i}  & \hbox{$ k \in (0,W_i-1), i \in (1,m)$} \\
        & P \{m,k | m,0\} = \frac{\Upsilon_C}{W_m}  & \hbox{$ k \in (0,W_m-1)$}. \\
        \end{array}
        }
      \right.
\end{equation}
where:
\begin{enumerate}[label=(\alph*)]
 \item The first equation signifies that the backoff time counter is decremented at the start of each slot. 
 \item The second equation states that, after a successful packet transmission, a new packet starts the backoff stage with a value of $0$. Thus, the value of the backoff counter is initially chosen uniformly in the range $(0,W_{0}-1)$. Unsuccessful transmissions are represented by the other equations. 
 \item In the third equation, when a transmission is unsuccessful, the backoff stage is incremented and the new initial value of backoff time counter is chosen uniformly in the range $(0,W_i)$. 
 \item The fourth equation signifies that the backoff stage is not increased once it reaches $m$.
\end{enumerate}
Let us consider that $b_{i,k} = \lim_{t\to\infty} P\{s(t) = i,b(t)=k\},\ i \in (0,m),\ k \in (0,W_i-1)$ represents the stationary distribution of the  Markov chain. Next, we compute a closed-form solution for the proposed Markov chain illustrated in Fig.~\ref{fig-model}. First, note that
\begin{align}
\label{eq2}
b_{i-1,0} \cdot \Upsilon_C = b_{i,0} \rightarrow b_{i,0}= \Upsilon_C^{m}b_{0,0} \ \ \ 0<i<m \nonumber\\
b_{m-1,0} \cdot \Upsilon_C = (1-\Upsilon_C)b_{m,0} \rightarrow b_{m,0}  = \frac{\Upsilon_C^{m}}{1-\Upsilon_C}b_{0,0}.
\end{align}
For each  $k \in (1,W_i - 1)$, by following the chain regularities, we derive the following set of equations
\begin{equation}
\label{eq3}
{b_{i,k}\ = \frac{W_i - k}{W_i} \cdot \
\left\{ \begin{array}{ll}
           (1-\Upsilon_C) \sum_{j=0}^{m} b_{j,0} & \hbox{$i=0$} \\
           \Upsilon_C\cdot b_{i-1,0} & \hbox{$0<i<m$}\\
           \Upsilon_C\cdot (b_{m-1,0} + b_{m,0}) & \hbox{$i =m$}.\\
         \end{array}
      \right.}
\end{equation}
On the basis of~(\ref{eq2}) and by using the fact that $ \sum_{i=1}^m b_{i,0} = \frac{b_{0,0}}{(1-\Upsilon_C)}$,~(\ref{eq3}) can be rewritten as
\begin{equation}
\label{eq4}
\begin{array}{lcl}
b_{i,k} = \frac{W_i-k}{W_i}b_{i,0} & \hbox{$i \in (0,m),$} & k \in (0,W_i-1).
\end{array}
\end{equation}
Therefore, based on~(\ref{eq2}) and~(\ref{eq4}), all the values of $b_{i,k}$ are expressed as functions of $b_{0,0}$ and of the collision probability $\Upsilon_C$. Finally, $b_{0,0}$ is determined by using the normalization condition and is simplified as
\begin{align}
\label{eq55}
\sum_{i=0}^m \sum_{k=0}^{W_i-1}b_{i,k} = \sum_{i=0}^m b_{i,0} \sum_{k=0}^{W_i-1}\frac{W_i-k}{W_i} \nonumber
\end{align}
\begin{equation}
\label{eq5}
=\frac{b_{0,0}}{2}\left[W \left(\sum_{i=1}^{m-1} (2\Upsilon_C)^i +\frac{(2\Upsilon_C)^{m}}{1-\Upsilon_C} \right) +\frac{1}{1-\Upsilon_C}\right]=1. 
\end{equation}
From~(\ref{eq5}), we can compute $b_{0,0}$ as
\begin{equation}
\label{eq6}
{
b_{0,0}= \frac{2(1-2\Upsilon_C)(1-\Upsilon_C)}{(1-2\Upsilon_C)(W+1)+\Upsilon_C W(1-(2\Upsilon_C)^m)}.
}
\end{equation}
Assume that $\nu$ is the probability of transmission in a randomly chosen slot. Since any transmission can occur while the back-off counter reaches $0$ regardless of the value of the back-off stage, we can define $\nu$ as   
\begin{align}
\label{eq7}
\nu & = \sum_{i=0}^m b_{i,0} = \frac{b_{0,0}}{1-\Upsilon_C} \nonumber \\
     & = \frac{2(1-2\Upsilon_C)}{(1-2\Upsilon_C)(W+1)+\Upsilon_C W(1-(2\Upsilon_C)^m)}.
\end{align}

\subsubsection{Throughput Estimation} 

Let $\mathcal{T}$ denote the normalized system throughput, expressed as the fraction of time duration where the wireless channel is used to transmit payload bits successfully. To compute $\mathcal{T}$, we consider a randomly chosen slot. In a given slot time, assume that $P_{one}$ is the probability of at least one transmission. As $n$ stations are contending on the channel and every station can transmit with a probability $\nu$, we can define $P_{one}$ as
\begin{equation}\label{Pone}
P_{one} = 1 - (1 - \nu)^n.
\end{equation} 
Let us consider that $P_s$ is the probability of successful transmission on the selected channel given the probability of at least one transmission on the channel. In a time slot, the probability that $(n-1)$ stations are not transmitting is $(1-\nu)^{n-1}$. Therefore, $P_s$ is obtained as
\begin{equation}\label{Ps}
P_s = \frac{n \nu(1 - \nu)^{n-1}}{P_{one}} = \frac{n \nu(1 - \nu)^{n-1}}{1 - (1 - \nu)^n}.
\end{equation} 

By following the model in~\cite{bianchi2000performance}, we compute the system throughput as follows. Let the average size of packet payload be $E[P]$. In a slot, a transmission occurs successfully with probability $P_{one}P_s$. Thus, the average size of payload information which is transmitted in a slot is $P_{one}P_sE[P]$. The probabilities of empty slot, successful transmission, and collision in a slot are $(1-P_{one})$, $P_{one}P_s$, and $P_{one}(1-P_s)$, respectively. Therefore, $\mathcal{T}$ can be computed as
\begin{equation}\label{thr}
\mathcal{T} = \frac{P_{one}P_{s}E[P]}{(1-P_{one})\sigma +P_{one}P_{s}T_{s}+P_{one}(1-P_{s})T_{c}}.
\end{equation}
Here, $T_s$ and $T_c$ denote the average time when the channel is sensed busy due to a successful transmission and a collision, respectively. Moreover, $\sigma$ denotes the duration of an empty slot. Here, $T_s$, $T_c$, $E[P]$, and $\sigma$ are measured with the same unit. We assume that all packets have the same fixed payload size. By following the basic IEEE 802.11 channel access method~\cite{standardDot11} that applies {\em Short Interframe Space (SIFS)} and {\em DCF Interframe Space (DIFS)}, $T_s$ and $T_c$ are obtained as
\begin{equation}
\label{basic}
     \left\{ {
        \begin{array}{l}
         T_s = H + E[P] + SIFS + \delta + ACK + DIFS + \delta  \\
         T_c = H + E[P] + DIFS + \delta.  \\
        \end{array}
        }
      \right.
\end{equation}
Here, $H$ is the packet header with $H=PHY_{head}+MAC_{head}$, where $PHY_{head}$ and $MAC_{head}$ denote the PHY and MAC layer packet headers, respectively. $\delta$ represents the propagation delay and $ACK$ is the acknowledgement frame.  
Next, we describe the TS-based mechanism that utilizes the estimated throughput $\mathcal{T}$ from our Markov chain model.

\subsection{Thompson Sampling based Learning in IBAC}
\label{sec:TS}

The proposed IBAC follows TS~\cite{Thompson:1933}-based reinforcement learning to generate the probability distributions of the success rate of using an action. In TS, the actions are selected in a manner such that the decisions must balance between the exploration of new information to improve future performance and the exploitation of the past knowledge to maximize next performance. Consequently, without any prior knowledge, a TS-based mechanism learns a new environment by exploring configurations whose performances have not evaluated so far. In addition, by the exploitation, the effective configurations can be identified and applied. Moreover, the application of both exploration and exploitation leads to a continuous learning of an unexplored environment and responding accordingly. 
As a result, TS-based mechanism becomes an online learning approach where an initial training is performed by applying the exploration repeatedly at the initial stage.




\subsubsection{Thompson Sampling}

TS is an online heuristic technique based on Bayesian inference for choosing the action that tries to maximize the expected reward on the basis of a randomly drawn belief. The TS-based online learning mechanism applies a probabilistic distribution approach by engaging the {\em reward} which is associated with each taken {\em action} (arm). The learner that uses TS has a tendency to gain knowledge (exploration) as well as optimize its decisions by exploiting the existing knowledge (exploitation). 

Let $\mathcal{X}$ be a set of contexts, $\mathcal{A}$ be a set of actions and $\mathcal{R}$ denote a set of rewards. In each round of taking an action, the learner obtains a context $x \in \mathcal{X}$ and applies an action $a \in \mathcal{A}$. Consequently, it gets a {\em reward} $r \in \mathcal{R}$. This reward follows a probability distribution that depends on the present context and the selected action. The objective of the learner is to choose the actions that lead to the maximization of the cumulative rewards. Let $\theta$ be the distribution of $r$ and $\mathcal{E}$ denote the past observation triplet, where $\mathcal{E}=\{(x;a;r)\}$. TS is associated with the prior, likelihood and posterior distributions. Let $P(\theta)$ denote the prior distribution on $\theta$. For a given $\theta$, $a$ and $x$, let the likelihood of $r$ be denoted by $P(r|\theta ,a,x)$. Let the posterior distribution of $\theta$ be represented by $P(\theta|\mathcal{E}) \propto P(\mathcal{E}|\theta) P(\theta)$, where $P(\mathcal{E}|\theta)$ is the likelihood function.
In TS, the learner plays the action $a^*\in\mathcal{A}$ by following the probability that maximizes the expected reward, and this probability is expressed as
\begin{equation}\label{TS}
 \int \Pi \Big[\mathbb{E}(r|\theta ,a^*,x) = \max\limits_{a^{\prime}} \mathbb{E}(r|\theta ,a^{\prime},x)\Big] P(\theta|\mathcal{E}) d\theta.
\end{equation}
Here, $\Pi$ represents the {\em indicator function} and $\mathbb{E}$ signifies the expected value. 


Next, we present a detailed description of the proposed TS-based mechanism in IBAC. 

\subsubsection{Working Principle of IBAC}

Our Markov chain model explained in Section~\ref{sec:Markov_model} provides the estimation of the system throughput $\mathcal{T}$ in~(\ref{thr}). Then, on the basis of $\mathcal{T}$, the proposed TS-based mechanism maximizes the channel bonding level in DBCA in order to avoid the OWRP in the network. Let $x$ represent the measured SNR value of the channel. The SNR is dependent on both the signal strength and noise level. Here, noise denotes both noise and interference. Since higher SNR values can minimize the interference in a wider channel, the SNR is considered as the context $\mathcal{X}$ in the IBAC.
Different components of the IBAC are described in what follows.

\textbf{SNR buckets:}
An {\em SNR bucket} is a set of SNR values which have been measured for a channel. We consider a sequence of such SNR buckets in our system. Let $\mathcal{B}=\langle b_1, b_2, b_3, ..., b_\ell\rangle$ denote a sequence of SNR buckets such that $b_1 < b_2 < b_3 < ... <b_\ell$, where $b_j$ ($1\le j \le \ell$) is the $j^{th}$ SNR bucket and $\ell$ represents the total number of SNR buckets. Moreover, assume that the range of each SNR bucket is $d$. For instance, if $d=5$ and $b_3$ starts from $26$ dB, we have $26$ dB $\leq b_3\leq$ $30$ dB and $b_4$ starts from $31 dB$. Let us consider that $SNR_{max}$ and $SNR_{min}$ are the maximum and minimum SNR values of the channel, respectively. Hence, $b_1$ begins from $SNR_{min}$ and $b_\ell$ ends with $SNR_{max}$. All intermediate SNR values of the channel are distributed through $b_1$ to $b_\ell$ according to the SNR bucket range $d$. Let $S_t$ be the SNR of the channel at time $t$. After measuring the channel's SNR value, it is stored into the appropriate SNR bucket in $\mathcal{B}$ according to the bucket range $d$.

\textbf{Statistical Table:}
We design a statistical table, denoted by $\mathcal{L}$, to store the SNR values and the channel bonding levels selected under these SNR values. $\mathcal{L}$ is represented as $\mathcal{L}=\{\mathcal{B},\mathcal{M}\}$. 

\subsubsection{Probability Distributions}

In the IBAC, the action $a$ denotes selecting a channel bonding level in DBCA. At any time instant, let us consider that $\theta$ denotes the {\em channel bonding level} to be chosen for the next transmission by using DBCA and thus, $\theta \in \mathcal{M}$. We consider the estimated system throughput $\mathcal{T}$ as $\mathcal{E}$ in our system. Since low signal interference can lead to high throughputs in the network, the estimated system throughput $\mathcal{T}$ can be used as the basis of the estimation of the channel bonding level. Detailed descriptions of these parameters are given in what follows.

\begin{itemize}
 \item \textbf{Prior distribution:} The prior is computed for the channel bonding level to be selected next based on the present value of the SNR. Basically, prior determines the probability of estimating a channel bonding level $\theta \in \mathcal{M}$. The prior is identified by $P(\theta)$. At time $t$, let the present SNR value $S_t \in b_j$ and $b_j \subset \mathcal{B}_s$ ($1\le j \le \ell$), where $\mathcal{B}_s \subset \mathcal{B}$. Let $\theta_{b_j}$ be a channel bonding level previously selected under $b_j$, where $\theta_{b_j} \in \mathcal{M}_s$ and $\mathcal{M}_s \subset \mathcal{M}$. Now, let $\mathcal{L}_s=\{\mathcal{B}_s,\mathcal{M}_s\} \subset \mathcal{L}$. Therefore, $P(\theta)=P(\theta_{b_j})$ denotes the probability of occurences of $\theta_{b_j}$ in $\mathcal{L}_s$. In this way, $P(\theta)$ is computed for every channel bonding level which was selected under the SNR bucket $b_j$. 

 \item \textbf{Likelihood:} In our model, the likelihood is denoted by $P(\mathcal{T}|\theta)$. It plays an important role in the estimation of the system throughput $\mathcal{T}$ of the selected level of channel bonding $\theta \in \mathcal{M}$. Let $\mathcal{T}_{\theta_{b_j}}$ be the estimated system throughput under the channel bonding level $\theta_{b_j}$, and thus $P(\mathcal{T}|\theta)=P(\mathcal{T}_{\theta_{b_j}}|\theta_{b_j})$. Now, for the given SNR bucket $b_j \subset \mathcal{B}_s$, $P(\mathcal{T}_{\theta_{b_j}}|\theta_{b_j})$ represents the probability of $\mathcal{T}_{\theta_{b_j}}$ under the channel bonding level $\theta_{b_j}$. The likelihood is computed for each channel bonding level selected under the SNR bucket $b_j$.  

 \item \textbf{Posterior distribution:} Let us consider that $P(\theta|\mathcal{T})$ represents the posterior distribution in the IBAC. The posterior distribution computes the probability of selecting the best possible channel bonding level based on the prior distribution and the likelihood for the next round of channel bonding in DBCA. Hence, following TS, $P(\theta|\mathcal{T})$ is calculated as
$$P(\theta|\mathcal{T}) \propto P(\mathcal{T}|\theta) \times P(\theta)$$
\begin{equation}\label{proposed}
 \implies P(\theta|\mathcal{T}) = \mathcal{K} P(\mathcal{T}|\theta) \times P(\theta). 
\end{equation}

Here, $\mathcal{K}$ is a constant of proportionality where we consider the maximum level of channel bonding as the value of this constant, i.e., $\mathcal{K}=u$. 
The posterior distribution is computed for all the channel bonding levels which are considered in calculating the prior and the likelihood. Then, the channel bonding level which provides the maximum value of the posterior distribution is selected for the next turn of DBCA. Algorithm~\ref{algo1} describes the execution steps of the IBAC mechanism.
\end{itemize}

\begin{algorithm}[!t]
\caption{IBAC -- Algorithmic Description}
\label{algo1}
\begin{algorithmic}[1]
\scriptsize
\State \textbf{Initialization:} Let $t_{init}$ be the number of rounds used for the {\em exploration} of several channel bonding levels and populating the initial $\mathcal{L}$. For $t=1, 2, 3,...,t_{init}$, calculate the present SNR $S_{t}$ of the channel and choose a level of channel bonding $\theta$ uniformly at random in $\mathcal{M}$.
\While{$t>t_{init}$}
\State Measure the present SNR $S_{t}$ of the channel.
\State Identify the SNR bucket $b_j$ such that $S_{t} \in b_j$ ($1\le j \le \ell$).
   \If{$b_j \subset \mathcal{B}_s$ such that $\mathcal{L}_s=\{\mathcal{B}_s,\mathcal{M}_s\} \subset \mathcal{L}$}
      \State Compute $P(\theta)$ for each channel bonding level that belongs to $\mathcal{M}_s$.
      \State Calculate $P(\mathcal{T}|\theta)$ based on the estimated system throughput, for each channel bonding level that belongs to $\mathcal{M}_s$.
      \State Compute $P(\theta|\mathcal{T})$ by following~(\ref{proposed}).
      \State Select the channel bonding level $\theta$ that provides the maximum value of $P(\theta|\mathcal{T})$.
   \Else
      \State Choose a channel bonding level $\theta$ uniformly at random in $\mathcal{M}$.
   \EndIf
\State Update $\mathcal{L}$.
\EndWhile
\end{algorithmic}
\end{algorithm}

\subsubsection{Time Bound Analysis}

Since the proposed mechanism is based on Step-1 and Step-2, we present time bounds for Step-1 and Step-2 separately, and compute the time complexity of the proposed scheme by combining these time bounds. The descriptions of the time bound analysis of Step-1 and Step-2 are provided in what follows. 

\noindent{\textbf{Time bound of Step-$1$:}} Since the output of Step-$1$ is the normalized system throughput, we calculate the time bound of estimating the throughput. As given in Eqn.~(\ref{thr}), the time bound of calculating the normalized throughput $\mathcal{T}$ primarily depends on the computations of $P_{one}$ and $P_s$. Both $P_{one}$ and $P_s$ are influenced by the number of stations contending on the channel. Therefore, from Eqn.~(\ref{Pone}), the time complexity of $P_{one}$ is $T(P_{one})=\mathcal{O}(n)$, and from Eqn.~(\ref{Ps}), the time complexity of $P_s$ can be represented as $T(P_s)=\mathcal{O}(n)$. Now, from Eqn.~(\ref{thr}), it is noted that $\mathcal{T}$ is defined by a multiplicative factor of $P_{one}$ and $P_s$. Let the time bound of calculating $\mathcal{T}$ be $T(\mathcal{T})$. Thus, $T(\mathcal{T})$ is represented as
\begin{equation}\label{throughput}
 T(\mathcal{T})=\mathcal{O}(n\times n)=\mathcal{O}(n^2). 
\end{equation}

\noindent{\textbf{Time bound of Step-$2$:}} In Step-$2$, the TS mechanism is based on the prior, likelihood, and posterior computations. Since the prior $P(\theta)$ denotes the probability of occurences of $\theta_{b_j}$ in $\mathcal{L}_s$, where $b_j \subset \mathcal{B}_s$ and $1\le j \le \ell$, we need to use a searching mechanism in $\mathcal{L}_s$, where the time bound of the search is $\mathcal{O}(|\mathcal{L}_s|)$. For the given SNR bucket $b_j \subset \mathcal{B}_s$, since the likelihood $P(\mathcal{T}_{\theta_{b_j}}|\theta_{b_j})$ represents the probability of $\mathcal{T}_{\theta_{b_j}}$ under the channel bonding level $\theta_{b_j}$, the likelihood is computed for each channel bonding level selected under the SNR bucket $b_j$. In this case, a searching technique is also required to search $b_j$, and thus the time bound of the likelihood is $\mathcal{O}(|b_j|)$. For the posterior distribution, as it is computed for all the channel bonding levels which are considered in calculating the prior and the likelihood, the time bound of the posterior distribution calculation depends on the number of the channel bonding levels used in the prior and likelihood. Since the maximum level of channel bonding is $u$, the upper bound of the computation of the posterior distribution is $\mathcal{O}(u)$. Therefore, the sum of the time bounds of the prior, likelihood, and posterior distributions is $\mathcal{O}(|\mathcal{L}_s|+|b_j|+u)$. If the searching mechanisms are applied in sorted tables, the time bound of TS can be reduced to $\mathcal{O}(\log|\mathcal{L}_s|+\log|b_j|+u)$.

Moreover, in order to analyze the time complexity of the proposed IBAC, both exploration and exploitation need to be considered. 
In the exploitation, the search will require the highest time when the present SNR value is not in $\mathcal{L}$. Thus, the searching time is bounded by $\mathcal{O}(|\mathcal{L}|)$. If $\mathcal{L}$ is arranged in sorted order of the SNR values, the search time can be reduced to $\mathcal{O}(\log|\mathcal{L}|)$. However, in the exploration, the channel bonding level is chosen randomly, and thus the time complexity of the exploration is independent of the size of $\mathcal{L}$. Let the time bound of TS be $T(\text{TS})$. Therefore, considering the exploration and exploitation, the time complexity of the TS-based approach can be defined as
\begin{align}\label{TSbound}
T(\text{TS}) &= \mathcal{O}(\log|\mathcal{L}_s|+\log|b_j|+u+\log|\mathcal{L}|)\nonumber\\
 &= \mathcal{O}(\log|\mathcal{L}|+\log|b_j|+u).
\end{align}

\noindent{\textbf{Time bound of IBAC:}} 
Considering Eqns.~(\ref{throughput}) and~(\ref{TSbound}), the time bound of the IBAC can be represented as 
\begin{align}
T(\text{IBAC}) &= T(\mathcal{T})+T(\text{TS})\nonumber\\
  &= \mathcal{O}(n^2+\log|\mathcal{L}|+\log|b_j|+u).
\end{align}


\section{Performance Analysis}
\label{sec:performance}

The IBAC is implemented in NS-3.33~\cite{ns3} and its performance is analyzed through an infrastructure network based on IEEE 802.11ax standard with two APs and multiple wireless stations per AP. Therefore, each AP forms a BSS. The stations are contending for the channel and initiating data transmissions among themselves. The IBAC is implemented in the APs i.e., the APs run the algorithm. We evaluate the efficiency of the IBAC with respect to DBS~\cite{DBS:2018}, WiderCast~\cite{park2020widercast} and the standard approach of DBCA~\cite{standardDot11ac}. We refer to this standard approach as ``General''. In the simulation, we measure the TCP throughput and TCP traffic in both directions i.e., data transmissions from APs to stations and vice versa. Thus, both downlink and uplink traffic are considered. Therefore, hidden node related collisions are included in the simulation. 
Since the RTS/CTS scheme cannot detect the OWRP properly, the RTS/CTS threshold is disabled in the simulation. 
Unless specified otherwise, in order to capture multi-hop effects, the number of wireless stations per BSS is $20$. 
Details on the simulation setup are presented in Table~\ref{table:sim}. 


\begin{table}[!t]
\caption{PHY/MAC and Control Parameters Used in Simulation}
\scriptsize
\centering
\begin{tabular}{|p{3.7cm}|p{4.1cm}|}
\hline
\textbf{Parameter} & \textbf{Value}\\
\hline 
\hline
High throughput standard & IEEE 802.11ax\\
\hline
Type of WLAN & Infrastructure network\\
\hline
Channel bandwidth & $20/40/80/160$ MHz\\
\hline
Guard interval & $800$ ns\\
\hline
MIMO antenna & $3$\\
\hline
Traffic source & TCP traffic\\
\hline
TCP payload & 1448 Bytes\\
\hline
Congestion protocol & $TcpWestwood$\\
\hline
Data and control mode & Constant rate wifi manager \\
\hline
Frame aggregation & Disabled\\
\hline
Block acknowledgement & Disabled\\
\hline
Maximum physical data rate & $1201$ Mbps\\
\hline
Path loss model & Log-normal path loss model (path loss exponent=$3.0$)\\
\hline
Propagation delay model & Constant speed propagation delay model\\
\hline
Simulation time & $10$ seconds\\
\hline
Number of repetitions of simulation for each number of station, for computing average performance  & $10$ \\
\hline
Mobility model & Random walk 2d mobility model (``Mode: Time'', ``Time: 2s'',``Bounds: Rectangle (0, 200, 0, 200)'', ``Speed: UniformRandomVariable[Min=2.0, Max=4.0]'', ``Direction: UniformRandomVariable[Min=0.0, Max=6.283184]'')\\
\hline
Frequency band & 5 GHz\\
\hline
Number of AP & $2$\\
\hline
Number of stations & $2$ -- $30$\\
\hline
Range of SNR values & $25$dB -- $50$dB\\
\hline
Traffic direction & AP to stations and vice versa\\
\hline
RtsCtsThreshold & disable\\
\hline
FragmentationThreshold & disable\\
\hline
BerThreshold & $0.00001$ (default)\\
\hline
IsLowLatency & false\\
\hline
DefaultTxPowerLevel & $10$\\
\hline
\end{tabular} 
\label{table:sim}
\end{table}

\subsection{Baselines}

WiderCast~\cite{park2020widercast} allows the AP to dynamically adapt the channel width for multicast transmission. The primary idea is to increase the channel width to the widest permissible transmission bandwidth that can be available across all stations (clients) in a multicast transmission group. Supportable bandwidth information is gathered by the AP from all stations before each multicast data transmission begins. To minimize the time overhead required to collect the bandwidth information, a bandwidth signaling mechanism is proposed in~\cite{park2020widercast}, that allows multiple stations to simultaneously provide their supportable channel width details to the AP in the frequency domain. DBS~\cite{DBS:2018} designs a maximization problem to avoid the OWRP and employs both virtual and physical carrier-sense approaches. Stations within the receiver's TR or the sender's CR can recognize the signal and postpone their data transfers by setting the medium to busy. The RTS/CTS packets are transmitted only at the primary channel. After exchanging the RTS/CTS, the transmitter selects the channel bandwidth, modulation and coding scheme by solving the maximization problem to avoid the OWRP and CSDP and initiate the data transmission. The ``General'' is the standard approach of DBCA defined in IEEE 802.11ac network standard~\cite{standardDot11ac}.

\subsection{Analysis of Throughput}

\begin{figure}[!t]
\centering
\includegraphics[width=\linewidth]{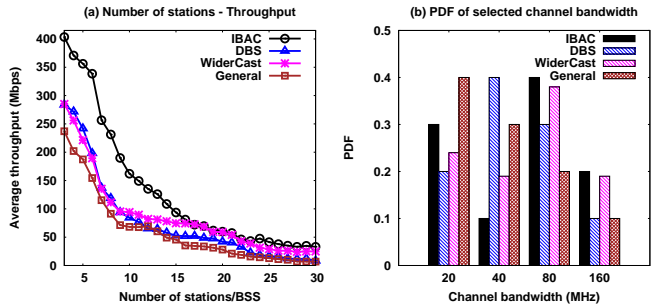}
\caption{Performance in terms of (a) average throughput and (b) PDF of selected channel bandwidth.}
\label{fig:thr-plr}
\end{figure}

Fig.~\ref{fig:thr-plr}(a) shows the performance comparison in terms of the average throughput which presents the per-user average throughput. To calculate it, total throughput (number of bytes transferred and received per second) at all stations is computed, and total throughput is then divided by the number of stations.
From Fig.~\ref{fig:thr-plr}(a), it can be observed that, for lower numbers of stations per BSS, the IBAC average throughput is significantly higher than that of the DBS, WiderCast and ``General'' approaches. When the number of stations per BSS is $5$, the IBAC has approximately $47\%$, $61\%$ and $90\%$ higher average throughputs than the DBS, WiderCast and the general DBCA schemes, respectively.
Due to the online learning-based approach, the proposed IBAC always adaptively chooses the channel width in such a way that the estimated system throughput is maximized. When the number of stations is low, the probability of the OWRP occurring is low, but it is not null. In this scenario, IBAC also intelligently adapts the channel width, resulting in increased network performance. Whereas, baselines are not learning-based schemes to dynamically adapt the channel width, and therefore when the number of stations is low, the performances of baseline approaches are still lower than IBAC.
As the number of stations increases, the collisions increase. However, given the proposed intelligent channel width selection, the IBAC can still minimize the OWRP, therefore achieving high throughputs compared to the other competing mechanisms. In this scenario, the IBAC shows better results because it exploits the past knowledge of the executions, which helps adjust with a network condition experienced earlier. 


The TS-based approach adapts the level of channel bonding by considering both the past and future transmissions. As a result, the IBAC always achieves a significantly higher average throughput than DBS, WiderCast and the general approach. DBS is a threshold-based scheme, and therefore it cannot well avoid the OWRP in varying network scenarios. WiderCast is based on a heuristic approach and does not consider the OWRP. Thus, it cannot use the secondary channel intelligently to minimize the impact of the OWRP after channel bonding. On the other hand, the general DBCA scheme does not handle OWRP in the network. 
As illustrated in Fig.~\ref{fig:thr-plr}(b) showing the probability density function (PDF) of the channel bandwidths selected, the IBAC always has a tendency to acquire a larger channel. We consider $20$ stations per BSS in Fig.~\ref{fig:thr-plr}(b) to include multi-hop analysis. This figure indicates that the probabilities of using the $80$ MHz and $160$ MHz channels are approximately $1.3$ and $2$ times higher in the IBAC than in DBS, respectively. Whereas, for both the $80$ MHz and $160$ MHz channels, these probabilities are moderately higher ($1.1$ times) in the IBAC compared to the WiderCast. The probabilities of using both the $80$ MHz and $160$ MHz channels are approximately $2$ times higher in the IBAC than in the ``General'' scheme. As a result, the average throughput is increased in the IBAC compared to the other competing mechanisms.  
From Fig.~\ref{fig:thr-plr}(a), for $25$ stations per BSS, the average throughput values are $23.02\%$, $42.39\%$ and $112.84\%$ higher in the IBAC than in WiderCast, DBS and the general DBCA approach, respectively. 

\subsection{Analysis of Packet Loss Rate}

\begin{figure}[!t]
\centering
\includegraphics[width=\linewidth]{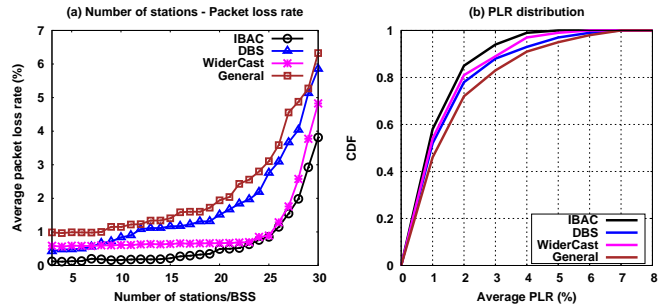}
\caption{(a) Average packet loss rate and (b) average packet loss rate distribution.}
\label{fig:cb-plr}
\end{figure}

Fig.~\ref{fig:cb-plr}(a) shows the performance comparison in terms of packet loss rate (PLR). The PLR is defined as the ratio of the number of lost packets to the number of packets transmitted, given as
$PLR=\frac{N_{tx}-N_{rx}}{N_{tx}}\times 100.$ Here, $N_{tx}$ and $N_{rx}$ define the total number of packets transmitted and received, respectively.
In the IBAC, the proposed Markov chain model for DBCA avoids the OWRP, and thus the packet loss rate is reduced. Furthermore, the TS-based mechanism considers both the past and future, which implies that the channel bonding level is selected such that the packet loss rate in the network is minimized.
On the other hand, the DBS can suffer from high packet loss due to the OWRP since it is a threshold-based approach. WiderCast does not address the OWRP issue, and thus WiderCast provides a higher packet loss rate when the OWRP arises due to the selection of a higher channel width. As a result, the IBAC provides a lower packet loss rate than the other competing schemes as illustrated in Fig.~\ref{fig:cb-plr}(a). This figure indicates that the average PLR values in the IBAC are approximately $68\%$, $26\%$ and $74\%$ lower than the DBS, WiderCast and general DBCA approach, respectively. Consequently, from Fig.~\ref{fig:cb-plr}(b), it can be observed that the cumulative density functions (CDFs) of the average PLR is lower in the IBAC than the other competing schemes.


\subsection{Analysis of Packet Loss Rate under Different Distances Between two APs}

\begin{figure}[!t]
\centering
\includegraphics[width=\linewidth]{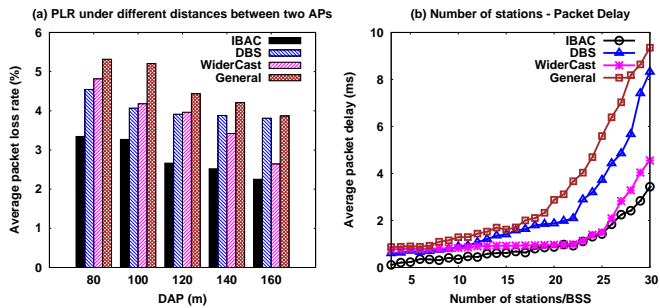}
\caption{Performance in terms of (a) average packet loss rate under different values of DAP and (b) average packet delay.}
\label{fig:plr-delay}
\vspace{-5mm}
\end{figure}

Since it was shown in Fig.~\ref{fig:owrp} that the distance between two APs can significantly influence the OWRP, we consider different distances between the two APs in our experiment. We consider a set of $d$ values of {\em distances between APs (DAP)}, where $d=[80,100,120,140,160]$.
Thus, the minimum and maximum distances between two APs are set to $80$m and $160$m, respectively. The intermediate values of DAP are chosen as $100$m, $120$m, and $140$m. For each BSS, there are $20$ stations with randomly selected positions. Fig.~\ref{fig:plr-delay}(a) shows the variations in the PLR under different distances between two APs. From this figure, it can be noted that, for the considered DAP range, the IBAC is associated to a lower PLR than the other competing schemes. For instance, for a DAP of $80$m, the average PLR of the IBAC is approximately $26\%$, $31\%$ and $37\%$ lower than the DBS, WiderCast and the general DBCA approach, respectively. This is because the probability of covering an AP by the IR of a station is significantly reduced for a given DAP in the IBAC compared to the other approaches. This reduction is due to the avoidance of the OWRP in our model. Consequently, the effect of collisions due to the OWRP is alleviated. In addition, the adaptive selection of channel bonding level by TS helps to select the channel bandwidth that minimizes the OWRP for the corresponding DAP values. In this regard, the exploitation of the past knowledge leads to a selection of channel bonding level considering the present channel condition. Therefore, this further reduces the packet loss rate in the IBAC.  
Whereas, due the lack of dynamicity in DBS and WiderCast, the best possible channel bonding level is not always selected. This leads to high packet collisions and consequently a higher packet loss rate in the network. The simple general DBCA approach cannot handle the OWRP, and thus it always suffers from high packet loss compared to the other schemes, as shown in Fig.~\ref{fig:plr-delay}(a).

\subsection{Analysis of Packet Delay}

Fig.~\ref{fig:plr-delay}(b) illustrates the performance of the IBAC in terms of average packet delay. In addition to the fact that the IBAC achieves a significantly higher average throughput than DBS, WiderCast and the general DBCA approach (Fig.~\ref{fig:thr-plr}(a)), the overall average delay is lower in the IBAC. It avoids OWRP, leading to a reduction of the probability of packet collision. As a result, its delay in packet transmissions is less than the DBS, WiderCast and general DBCA approach. From Fig.~\ref{fig:plr-delay}(b), the IBAC has approximately $53\%$, $10\%$ and $70\%$ lower average packet delay than the DBS, WiderCast and ``General'' schemes, respectively. 

\subsection{Analysis of Throughput Fairness}

\begin{figure}[!t]
\centering
\includegraphics[width=\linewidth]{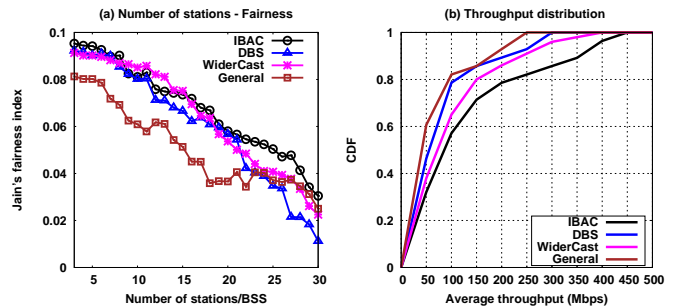}
\caption{Performance in terms of (a) fairness and (b) CDF of average throughput distribution.}
\label{fig:fairness-cdf}
\vspace{-5mm}
\end{figure}

As the number of stations increases, the congestion increases in the network, and consequently the probability of collisions increases. This reduces the fairness in the channel access and increases the IR in the network since all stations have a tendency to acquire large channel widths. As a result, the OWRP is increased in a congested network scenario. 
The analysis of the throughput's Jain's fairness index~\cite{HuaizhouFairness:2014} is presented in Fig.~\ref{fig:fairness-cdf}(a) which shows that the fairness gets better in the IBAC as the network becomes more congested. In a congested environment, after a few runs, the IBAC finds the best possible channel bonding level, which has been utilized most frequently in the past and has the highest probability of achieving the maximum throughput in the next transmission. In this way, the IBAC tries to adapt with different network scenarios. 
However, the DBS, WiderCast and general DBCA approach are not adaptive, and thus their performances in terms of fairness are poor in congested networks compared to the IBAC. 

\subsection{Analysis of Average Throughput Distribution}

Fig.~\ref{fig:fairness-cdf}(b) shows the CDF of the average throughput of the considered approaches. Since the average IBAC throughput is significantly higher than that of the other schemes, the CDFs of the DBS and general DBCA approach converge around $300$ Mbps, where the CDF values are high in the $50$ Mbps -- $250$ Mbps range. On the other hand, the CDF of the WiderCast converges around $350$ Mbps, whereas the density of the IBAC spans through $100$ Mbps -- $450$ Mbps. In the IBAC, the exploitation of the past execution knowledge enables high throughputs. In this regard, the channel bandwidth level is selected by considering the present channel condition such that throughput is enhanced. As a result, the overall average throughput distribution is higher in the IBAC than the other competing schemes. 

\subsection{Distribution of PLR Across Different Channel Widths}

\begin{figure}[!t]
\centering
\includegraphics[width=\linewidth]{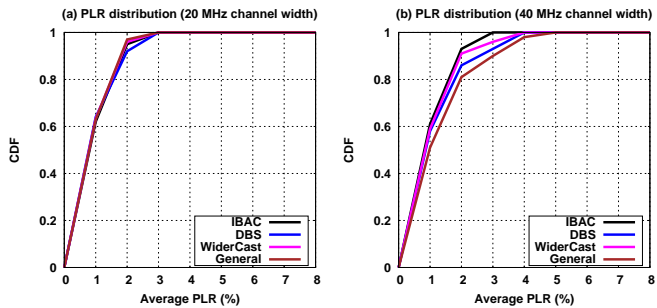}
\caption{Distribution of PLR across channel widths: (a) 20 MHz and (b) 40 MHz.}
\label{fig:CB2040}
\end{figure}

\begin{figure}[!t]
\centering
\includegraphics[width=\linewidth]{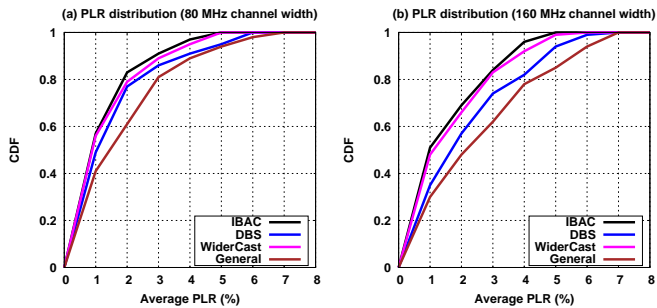}
\caption{Distribution of PLR across channel widths: (a) 80 MHz and (b) 160 MHz.}
\label{fig:CB80160}
\end{figure}

Fig.~\ref{fig:CB2040} and Fig.~\ref{fig:CB80160} illustrate the probability distributions of the average PLR across different channel widths. From Fig.~\ref{fig:CB2040}(a), it can be noted that the distribution of the PLR across the $20$ MHz channel is almost the same for the three mechanisms. This is because the $20$ MHz channel is less prone to interference due to the hidden channel problem. However, this problem is increased for higher channel widths; where as the channel width increases, the interference becomes higher. In the IBAC, when the present SNR of the channel is not found in the SNR bucket, channel widths are explored randomly from the available set of channel bandwidths. Therefore, all channel widths can be considered to calculate the prior, the likelihood, and the posterior distributions, in order to determine the channel bandwidth for the next data transmission. The channel width that has been used most frequently in the past gets the highest probability to be selected for the next transmission, such that the average throughput can be increased, and consequently the packet loss rate is decreased. This type of channel bandwidth selection leads to the selection of channel widths that produce low packet loss rates. 

From Fig.~\ref{fig:CB2040}(b) and Fig.~\ref{fig:CB80160}, it can be observed that, with the help of TS-based adaptive learning, the IBAC manages to provide less PLR compared to the other three schemes in scenarios of high channel bandwiths. Fig.~\ref{fig:CB2040}(b) indicates that, in the $40$ MHz channel width, the IBAC achieves approximately $7\%$, $5\%$ and $12\%$ reductions in the average PLR compared to the DBS, WiderCast and ``General'' schemes, respectively. 
The major savings come in the case of $160$ MHz channel, as can be noted from Fig.~\ref{fig:CB80160}(b). Precisely, across the $160$ MHz channel width, the PLR distribution of the IBAC is approximately $33\%$, $21\%$ and $43\%$ less than that of the DBS, WiderCast and ``General'' approaches, respectively.  

\subsection{Convergence Analysis}

\begin{figure}[!t]
\centering
\includegraphics[width=\linewidth]{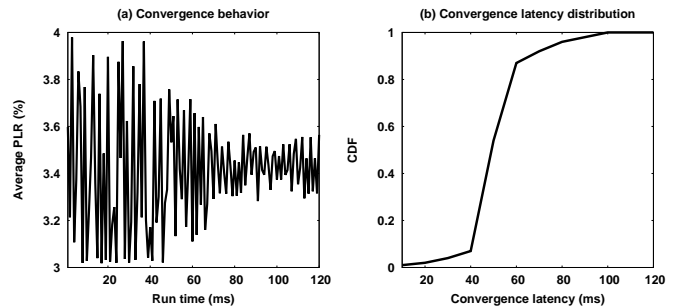}
\caption{(a) Convergence behavior and (b) distribution of convergence latency.}
\label{fig:conv}
\vspace{-5mm}
\end{figure}

Fig.~\ref{fig:conv}(a) illustrates the convergence behavior of the IBAC. In this case, we compute the average PLR and consider $30$ stations per BSS. 
In order to impose variable signal quality in the network, we randomly vary the SNR of the channel between $25$dB--$50$dB.
At the initial stage of the execution of the IBAC, the rate of exploration is very high to obtain knowledge about the environment where the system is run. Consequently, there is a high fluctuation in the average packet loss rate at the initial phase of the execution of IBAC. However, as time progresses, the statistical table is enriched with more information, which is used in exploiting the past knowledge. As a result, the average PLR becomes quite stable as the execution time increases, compared to the early phase of the execution. From Fig.~\ref{fig:conv}(a), it can be noted that the average PLR of the IBAC converges to a significantly low fluctuation at around $80$ms, which is quite low for the adaptive learning of a new wireless environment.

We compute the convergence latency by randomly changing the number of stations per BSS between $5$--$30$. 
The CDF of the convergence latency is shown in Fig.~\ref{fig:conv}(b). From this figure, it can be observed that the distribution of the convergence latency is high in the $40$ms--$90$ms range, which is fast even in networks with rapidly varying channel conditions and station density. A dynamic network scenario highly impacts on the channel condition. In the IBAC, the past knowledge is exploited in the learning and if it observes a similar network condition in the past, the convergence is achieved quickly. Otherwise, in the exploration, the learning converges when it observes an optimal solution.

\subsection{Performance of IBAC Considering Channel Aggregation}

\begin{figure}[!t]
\centering
\includegraphics[width=\linewidth]{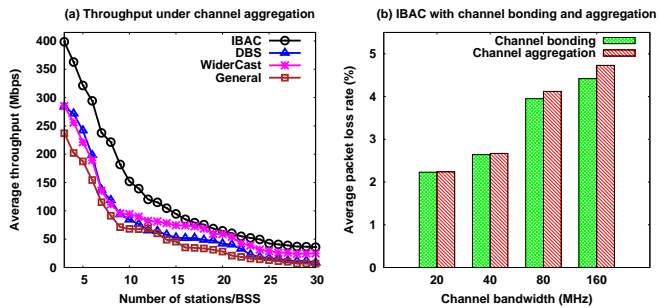}
\caption{(a) Average throughput considering channel aggregation and (b) average PLR of IBAC considering channel bonding and channel aggregation.}
\label{fig:ca}
\end{figure}

In order to analyze the performance of the IBAC while applying channel aggregation, we compute the average throughput and PLR, as shown in Fig.~\ref{fig:ca}.

\subsubsection{Average Throughput}
From Figs.~\ref{fig:thr-plr}(a) and~\ref{fig:ca}(a), it is noted that the average throughput of the IBAC is improved when channel aggregation is considered. As the network becomes congested, the number of available contiguous channels decreases. Specifically since aggregated channels may be non-contiguous in channel aggregation, channel aggregation helps create a higher channel bandwidth for the DBCA. As a result, in a dense network, the DBCA probability is higher in channel aggregation than in channel bonding, and consequently the average throughput is improved with channel aggregation. For instance, when the number of stations is $20$, the IBAC with channel aggregation has an average throughput approximately $8.28\%$ higher than the IBAC with channel bonding. Overall, from Figs.~\ref{fig:thr-plr}(a) and~\ref{fig:ca}(a), it is noted that, by using channel bonding or channel aggregation, the IBAC can provide a higher average throughput than baseline mechanisms.

\subsubsection{Average PLR across Different Channel Widths}
Fig.~\ref{fig:ca}(b) shows the average PLR of the IBAC considering channel bonding and channel aggregation for different channel bandwidths. Since the aggregated channels may not be contiguous, the intermediate channels (between the aggregated channels) may be used by other stations, causing interference in the network. The interference increases as the number of aggregated channels increases, and consequently the average PLR becomes higher as the channel bandwidth increases. Therefore, as the channel width increases, channel aggregation provides higher packet loss than channel bonding. For instance, it is observed in Fig.~\ref{fig:ca}(b) that for a channel bandwidth of $160$ MHz, the average PLR is approximately $6\%$ higher in channel aggregation than in channel bonding.


\section{Conclusion}
\label{sec:concl}

Collisions due to the OWRP in multi-hop wireless networks can significantly degrade the performance of HE-WLANs. In this paper, the proposed IBAC addresses the OWRP in dynamic bandwidth channel access mechanism in HE-WLANs. We apply a probabilistic approach in selecting the channel bonding level. In this regard, we first design a two-dimensional Markov chain model for the DBCA mechanism to overcome the OWRP and estimate the system throughput in the network. On the basis of this estimation, we then propose an adaptive MAC layer mechanism to choose a channel bonding level during the DBCA approach. The IBAC is based on Thompson sampling, a Bayesian approach, for adapting the channel bonding level to the network condition. Simulation results of the proposed IBAC show that, compared to other competing mechanisms, the proposed mechanism can significantly improve the system performance by avoiding the OWRP. 
By avoiding the OWRP and intelligently adapting channel width, the IBAC can lead to the performance improvement of WLAN standards such as IEEE 802.11ax networks. 
Since the wireless network is a time-varying system, where the signal strength changes abruptly, the convergence time of the proposed scheme may vary in different network scenarios.

\section{Acknowledgement}
This work was supported by the Canada Research Chair Program tier-II entitled ``Towards a Novel and Intelligent Framework for the Next Generations of IoT Networks''.

\bibliographystyle{IEEEtran}
\bibliography{DBCA_reference}

\vfill 

\end{document}